# First-Principles Theory for Schottky Barrier Physics

Dmitry Skachkov, Shuang-Long Liu, Yan Wang,[*] Xiao-Guang Zhang,[†] Hai-Ping Cheng

*Center for Molecular Magnetic Quantum Materials, University of Florida, Gainesville, Florida 32611, USA*
*Department of Physics and the Quantum Theory Project, University of Florida, Gainesville, Florida 32611, USA*

We develop a first-principles theory for Schottky barrier physics. The Poisson equation is solved self-consistently with the electrostatic charge density over the entire barrier using the density functional theory (DFT) electronic structure converged locally, allowing computation of a Schottky barrier entirely from DFT involving thousands of atomic layers in the semiconductor. The induced charge in the bulk consists of conduction and valence band charges from doping and band bending, as well as charge from the evanescent states in the gap of the semiconductor. The Schottky barrier height is determined when the induced charge density and the induced electrostatic potential reach self-consistency. Tests on the GaAs – graphene and Si/Al heterostructures yield Schottky barrier height, width, along with depletion and inversion layers obtained self-consistently as functions of temperature and bulk doping.

DOI:

## I. INTRODUCTION

The physical property of a Schottky contact is one of the most important topics in semiconductor (SC) physics, because metal-semiconductor contacts are the main components of semiconductor electronic devices.[1,2,3,4,5,6,7,8] Schottky barriers form at the interface between a semiconductor and a metal, or between two dissimilar semiconductors. At such an interface, charge transfer between the two materials is necessary in order to bring their Fermi levels into alignment. However, in a semiconductor the Fermi level density of states is nominally zero, requiring a space charge layer of macroscopic thickness to accommodate the amount of charge transfer needed. A corresponding shift in the electrostatic potential produces a potential barrier for the carriers.[9,10] The physics of Schottky barrier is rich and has far-reaching impacts. It is important in wide ranging problems including how to make good contact between the semiconductor and the metal layer in order to avoid surface charge accumulation and produce high-quality Schottky contacts,[11,12] how to use gate dielectric to avoid Fermi level pinning[13,14] and reduce tunneling from metal to semiconductor.[15,16,17,18] A Schottky barrier is characterized by the barrier height and thickness, both of which depend on many factors, for example, the quality and structure of the semiconductor surface,[19,20,21,22] electronic structure of the surface,[6] and the lattice strain.[23]

The macroscopic size of a Schottky barrier typically exceeds thousands of atomic layers of the semiconductor, and makes its self-consistent computation from first-principles a prohibitively difficult task. Lacking a fully-self-consistent density functional theory (DFT) solution for the entire barrier region, DFT studies focused on the problem of band alignment,[10,24,25,26,27,28,29,30,31,32,33,34,35,36,37] which is usually determined by comparing characteristic energy levels or core electron levels at the interface to that in the bulk semiconductor,[38,39] or on averaging the electrostatic potential.[40,29] Such calculations typically require separate interface and bulk calculations, with no self-consistent connection between them. Some studies included a few atomic layers[41] of the bulk semiconductor in the interface calculation, but the electrostatic potential and in particular the electric field over these layers are far from the values in the bulk. Such calculations can provide, in principle, the value of the Schottky barrier height (SBH). However, they do not provide a self-consistent first-principles description of the electrostatic potential profile of the Schottky barrier. Specifically, none of the first-principles calculations is able to obtain the barrier thickness.

Many empirical models for Schottky barrier[3,5,42,43,44,45,46] were developed to solve the Poisson equation with the help of experimental parameters. Several models were developed in order to connect empirical models with DFT calculations, including, for example, models for solving the Poisson equation for the electrostatic potential,[19,47] and models using current-voltage characteristics.[48] With the help of these models it is possible to find the charge separation at the interface and the value of SBH. However, these models require empirical parameters derived from experiment, and in these models the electrostatic potential due to Schottky contact is not included self-consistently into DFT calculations. Moreover, different experimental techniques for Schottky barrier study, for example, based on *I-V* (*C-V*) curves,[13,49,50,51] or by infrared photoresponce measurements, give different results.[52] Inhomogeneous barriers of Schottky contacts play also important role in experimental finding of SBH.[46] All these effects give ambiguousness in experimental determination of Schottky contact parameters.

---
[*] Present address: Advanced Materials Lab, Samsung Research America, 665 Clyde Ave, Mountain View, CA 94043, USA
[†] email: xgz@ufl.edu

In the current work we present a DFT approach that enables the simultaneous calculations of all atomic layers from the interface to the bulk semiconductor over the distance of the entire Schottky barrier, for arbitrary barrier thickness (Schottky barrier width). The DFT calculations account for the shift in the electrostatic potential as a function of the distance to the interface as well as the local electric field which differ from layer to layer as a function of the distance from the surface. The electrostatic potential and the local electric field are solved from the Poisson equation for the entire semi-infinite system. Both the height and the thickness of the Schottky barrier, along with the charge redistribution over the entire semi-infinite system are obtained as the results of the self-consistent calculation.

The rest of the paper is organized as the following. In the Theory section we start from the Poisson equation and its boundary conditions. This is followed by a description of the metal-induced gap states (MIGS) in the semiconductor and the induced charges due to MIGS. The self-consistent procedure for calculating the electrostatic potential from the DFT is also presented. In the Computational Approach section we will present details of calculations. In the Result section we present results of Si/Al heterostructure, as well as GaAs-graphene system.

## II. THEORY

### A. Electrostatic potential from induced charges

The charge separation on Schottky contact creates an additional global electrostatic potential, which must be added to the electrostatic potential due to local electrons and nuclei calculated from the DFT. The Poisson equation for the entire system is,

$$\nabla^2 \tilde{V}(\mathbf{r}) = -\tilde{\rho}(\mathbf{r})/\varepsilon_0 \quad (1)$$

The charge density includes electrons and nuclei,

$$\tilde{\rho}(\mathbf{r}) = -e \sum_j |\psi_j(\mathbf{r} - \mathbf{R}_j)|^2 + e \sum_j Z_j \delta(\mathbf{r} - \mathbf{R}_j) \quad (2)$$

where $\mathbf{R}_j$ and $Z_j$ are positions and charges of the nuclei. Away from the interface region, the electronic structure is almost bulk-like, but has small and gradual variations that integrate to form the Schottky potential. To calculate such variations self-consistently, we discretize the Schottky barrier region into a number of layers. Within each layer, we separate the electrostatic potential into a part that is directly calculated from DFT for a bulk system that is a periodic repetition of the current layer, and a second part, which we call the Schottky potential and denote as $V(\mathbf{r})$, that accounts for all the excess charge responsible for the Schottky barrier. The first part is the electrostatic potential of a neutral system. The DFT calculation can include a local electric field (see below) but does not include any excess charge due to doping or MIGS.

The calculation of the Schottky potential $V(\mathbf{r})$ can be greatly simplified if we note that away from the charge source, the Poisson equation is

$$\nabla^2 V = 0.$$

The general solution within a layer-periodic geometry is

$$V(\mathbf{r}) = \sum_{mn} V_{mn} e^{i(mk_x x + n k_y y)} f_{mn}(z)$$

$$f_{mn}(z) = \begin{cases} z, & m=0, n=0 \\ e^{-\sqrt{m^2 k_x^2 + n^2 k_y^2}\, z}, & m \neq 0, n \neq 0 \end{cases}$$

where $k_x$ and $k_y$ are the two-dimensional reciprocal lattice vectors for a square lattice (generalization to non-square lattices is straightforward). It is clear that the part of the potential that oscillates in the $xy$ plane decays exponentially in the $z$ direction over the distance of the lattice constant. If we neglect the small oscillatory contribution from the excess charge within the current layer, we can consider the Schottky potential to be constant in $x$, $y$ within the layer, and varying only in $z$ direction (see Fig. 1).

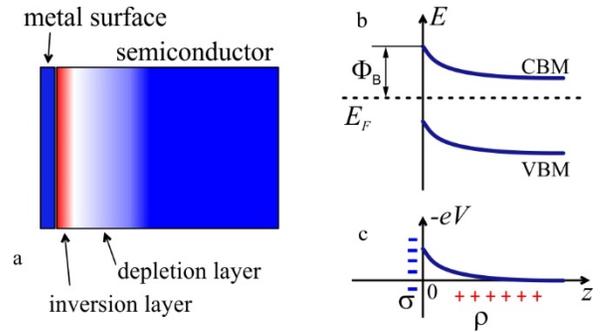

Figure 1: Schottky contact between metal and n-type semiconductor. (a) the charge redistribution in the bulk of *n*-type semiconductor (blue represents negative charge and red positive); (b) band bending in semiconductor; (c) potential energy of electron in Schottky potential for $\sigma < 0$. $\Phi_B$ shows the SBH. $\sigma$ is the surface charge on the metal surface, and $\rho(z)$ is the charge distribution in the semiconductor bulk creating the excess electrostatic potential $V(z)$. Migration of mobile electrons in *n*-type semiconductor into the metal surface creates the depletion layer near the semiconductor surface. Due to band bending thermally activated holes create the inversion layer.

The gradient of Schottky potential creates an electric field which acts on each layer that in turn induces an additional dipole response from the SC. The Schottky potential, which should be added to every cell of DFT calculations, can be found by solving the Poisson equation

$$\nabla^2 V(z) = -\frac{\rho(z)}{\varepsilon_0} + \frac{\nabla \cdot \mathbf{P}(z)}{\varepsilon_0} \qquad (3)$$

where $\rho(z)$ is the density of induced charges in the semiconductor due to doping, thermal holes and electrons, and MIGS (see Section II.B). The term $-(\nabla \cdot \mathbf{P})$ is the charge density due to the polarization $\mathbf{P}$ of the media due to local electric field arises from Schottky potential.

We set the potential at $z \to +\infty$ to zero (see Fig. 1),

$$V(z)\big|_{z \to +\infty} = 0 \qquad (4)$$

and the boundary condition at the surface corresponds to

$$\frac{\partial V}{\partial z}(0) = -\frac{\sigma}{\varepsilon_0} + \frac{P(0)}{\varepsilon_0} \qquad (5)$$

where $\sigma$ is the surface charge density accumulated at the contact surface and $P(0)$ is the polarization of the semiconductor at the surface.

The electrostatic potential due to charge $\rho(z)$ in the bulk and the charge at the metal surface can be calculated as potential from the charged planes $\rho dz'$ and $\sigma$:

$$V_\rho(z) = -\frac{1}{2\varepsilon_0}\int_0^{+\infty} \rho(z')|z-z'|dz' - \frac{1}{2\varepsilon_0}\sigma|z| \qquad (6)$$

where $\varepsilon_0$ is the vacuum dielectric constant. $\sigma$ is equal to the total charge in the bulk semiconductor with opposite sign

$$\sigma = -\int_0^{+\infty} \rho(z')dz' \qquad (7)$$

to ensure total charge neutrality of the system. The first-principles approach to calculate the excess charge density $\rho$ will be described in subsection B. The expression (6) can be simplified by taking into account the charge neutrality condition (7) and setting the potential at $z \to +\infty$ to zero according to the boundary condition (4),

$$V_\rho(z) = -\frac{1}{\varepsilon_0}\int_z^{+\infty} \rho(z')(z'-z)dz' \qquad (8)$$

Electric field due to the charge redistribution induces dipole moment of the media (polarization $\mathbf{P}$) which creates, in turn, the electric field with opposite direction. These induced dipole moments create an additional contribution to the electrostatic potential

$$V_P(z) = \frac{1}{2\varepsilon_0}\int_0^{+\infty} P(z')\text{sign}(z-z')dz' \qquad (9)$$

where $P(z)$ is the polarization of the bulk semiconductor in the direction of the field. Using the same procedure as with derivation of (8) we can rewrite (9) in the following form

$$V_P(z) = -\frac{1}{\varepsilon_0}\int_z^{+\infty} P(z')dz' \qquad (10)$$

The total excess electrostatic potential is the sum of the two contributions (8) and (10),

$$V(z) = V_\rho(z) + V_P(z) \qquad (11)$$

The electrostatic potential (11) has two impacts on the Schottky contact. First, it shifts the energy of electronic states for each layer of the bulk semiconductor by potential energy -$eV(z)$, and second, the gradient of the potential $V$ defines the local electric field acting on the current layer of the semiconductor. Previous attempts of first-principles methods for calculating Schottky barrier based on DFT[10-19] does not include the Schottky potential into the DFT self-consistency cycle due to very wide depletion layer (around 100-1000 Å for high doping concentration), yielding band structure of the semiconductor without the effect of the local electric field due to the Schottky contact.

### B. Induced charge distribution due to band bending

Bringing a semiconductor into contact with a metal over a clean interface will accumulate or deplete free mobile charges from the semiconductor, depending on the difference in the electrostatic potential on two sides of the interface. This redistribution of the charges creates a volume charge density $\rho(z)$ in the semiconductor region next to the interface and a surface charge density $\sigma$ on the metal side, keeping the total charge of the system equal to zero. The charge distribution near an interface on the semiconductor side usually contains two main contributions. One is from the so-called metal-induced gap states [53] or in the case of a surface, surface-induced gap states (SIGS), the other is from free carriers, including both holes and electrons in the semiconductor due to doping or thermal activation. The total excess charge density is the sum of the contributions from holes, electrons, MIGS, and the charged dopants that act as a compensating background charge in the bulk,

$$\rho(z) = \rho_h(z) + \rho_e(z) + \rho_{\text{MIGS}}(z) + \rho_d \qquad (12)$$

where the dopant charge is equal to $\rho_d = -\left[\rho_h(z) + \rho_e(z)\right]\big|_{z \to +\infty}$ and depends on the Fermi level position.

The charge distribution of the holes in the valence band is calculated according to the Fermi-Dirac distribution and increases (or decreases) closer to the interface due to the bending of the valence band by the electrostatic energy -$eV(z)$, where $V(z)$ is the electrostatic potential due to the induced excess charge given by (11). To account for this extra charge the density of states (DOS) of each layer of the bulk semiconductor is calculated under the influence of the electric field $\mathcal{E} = -\nabla V(z)$. To obtain the induced charge, integration over the energy up to valence band maximum (VBM) $E_{VBM} - eV(z)$ is equivalent to the shifting of the Fermi level down by $-eV(z)$:

$$\rho_h(z) = \frac{e}{\Omega} \int_{-\infty}^{E_{VBM}-eV(z)} [1-f(E)] D_b [E-(-eV(z))] dE \quad (13)$$

where $\Omega$ is a volume of the cell, $f(E)$ is the Fermi-Dirac function and $D_b[E]$ is the DOS of the semiconductor bands, modified by the local electric field due to the induced charge density.

The electron carrier contribution to the excess charge distribution is calculated by a similar integration of the DOS from the conduction band minimum (CBM) up to $+\infty$:

$$\rho_e(z) = -\frac{e}{\Omega} \int_{E_{CBM}-eV(z)}^{+\infty} f(E) D_b [E-(-eV(z))] dE \quad (14)$$

Another contribution from the electron density is from the MIGS or SIGS, which are the evanescent states within the band gap[3,5,53,54,55,56,57,58,59] and form on the surface of the semiconductor and penetrate into the bulk, generating a charge distribution. A powerful tool for calculating the MIGS from first-principles is the framework of complex band structure (CBS)[56,60,61,62] in semiconductor. This method has been used, for example, for calculation of carrier mobility in heterojunctions[63,64] and tunneling coefficients in strong-correlated systems.[65]

The wave functions of the Bloch electrons in the metal have exponential tails extending into the semiconductor, and manifest as the evanescent states. These evanescent states are described by the CBS[61,62] and are energy dependent. We express the evanescent state wave functions in the form

$$\Psi(z) \sim \chi(\mathbf{k}) e^{-\kappa_z(E,\mathbf{k})z} \quad (15)$$

where $\chi(\mathbf{k})$ is a periodic function in the plane perpendicular to $z$, $\kappa_z$ is the penetration (decay) rate, and the density of states of MIGS is defined by expression (16). It is necessary to take into account that due to Schottky contact each layer of the bulk semiconductor exhibits the influence of local electric field due to Schottky potential, and thus the CBS can be different from layer to layer of the semiconductor.

The MIGS contribution is proportional to the exponential factor of the CBS decay rate $\kappa_z(E,\mathbf{k})$:

$$\rho_{MIGS}(z) = -\frac{e}{\Omega} \int_{E_{VBM}-eV(z)}^{E_{CBM}-eV(z)} f(E) D_{MIGS}(\sigma, E-(-eV(z)), z) dE \quad (16)$$

where

$$D_{MIGS}(\sigma, E, z) = \int d\mathbf{k} \{D_i^L[\sigma,E,\mathbf{k}] \exp[-2\kappa_z^L(E,\mathbf{k})z] + D_i^H[\sigma,E,\mathbf{k}] \exp[-2\kappa_z^H(E,\mathbf{k})z]\} \quad (17)$$

is the DOS of a MIGS. $D_i^{L/H}[\sigma, E, \mathbf{k}]$ are the contributions to the projected density of states (PDOS) of the interfacial layer for light and heavy holes, correspondingly, and $\kappa_z^{L/H}(E,\mathbf{k})$ are the decaying rates for light/heavy holes. In (17) we also take into account that the density of states is proportional to the square of the wave function (15).

The semiconductor energy gap near the surface is shifted due to the Schottky potential, therefore the MIGS will emerge and disappear near the bended VBM and CBM (see Fig. 2). These states are taken into account by matching boundary conditions for $D_i^{L/H}[\sigma,E,\mathbf{k}]$ at each position $z$.

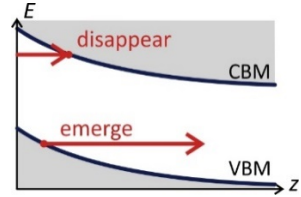

Figure 2: Emergence and disappearance of MIGS near the bended VBM and CBM at the gap of semiconductor which exposed to the electron wave functions in the metal according to (16).

The energy shift $\Delta E$ of the surface states due to migrated electrons to the surface (see Fig. 3) is determined by the surface charge density $\sigma$,

$$\sigma = -\frac{e}{S} \int_{E_F-\Delta E}^{E_F} f(E) D_0 (E-(-eV(0))) dE \quad (18)$$

where $(E_F - \Delta E)$ is the Fermi level of the neutral interface, $D_0(E)$ is the PDOS of the surface interface, which includes the metal surface and the surface layer of the semiconductor, and $S$ is the surface area of the cell.

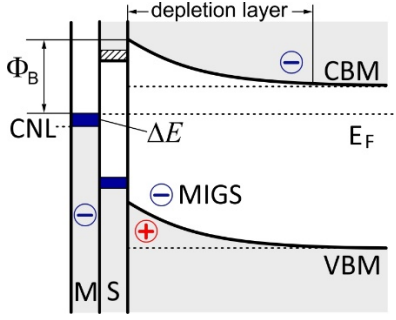

Figure 3: Scheme for *n*-type doped semiconductor in contact with metal surface (M), blue area $\Delta E$ shows energy filling level due to migrated electrons from semiconductor, minus sign on metal surface shows migrated electrons from semiconductor, minus sign in the gap of semiconductor shows MIGS states, and plus sign near the bended VBM shows thermally activated holes responsible for creation of inversion layer. CNL is a charge neutrality level corresponding to zero charge on the interface. On the semiconductor surface (S) it is shown two areas corresponding to the surface states, dashed area is unfilled and blue area is filled surface states. $\Phi_B$ shows the SBH.

### C. Self-consistent solution for Schottky contact

The effect of the electrostatic potential due to the Schottky contact can be added to the DFT calculations by introducing an equivalent electric field $\mathcal{E} = -\nabla V(z)$ which differs from layer to layer in the bulk semiconductor, in addition to the local shift of the Fermi energy relative to the band structure. The induced charge density due to the contact is calculated from the local electrostatic potential (11), and induced polarization due to electric field $\mathcal{E} = -\nabla V(z)$ is calculated in DFT. On the other hand, the overall shift in the electrostatic potential due to the Schottky contact is defined by the induced charge (12). Then the Schottky potential will be defined as self-consistent equilibrium between $V$, $\rho$, $\mathcal{E}$, $P$, and the charge on the interface $\sigma$:

$$V\big[\rho[V(z)], P(z), \mathcal{E}, \sigma\big] \equiv V(z) \quad (19)$$

In order to find the equilibrium (19) we use a self-consistent procedure. A good approximate initial charge distribution is,

$$\rho(z) = \rho_0\, e^{-z/z_0} \quad (20)$$

The corresponding charge accumulated at the contact surface due to (20) is

$$\sigma = -\rho_0 z_0 \quad (21)$$

The electrostatic potential from charge distribution (20) according to (8) is

$$V(z) = V_0\, e^{-z/z_0} \quad (22)$$

where $V_0 = \rho_0 z_0^2 / \varepsilon_0$, $\rho_0$ and $z_0$ are taken from the experimental data,[66],[67] which yield starting estimates $z_0 \sim 1{,}000$ Å and $-eV(0) = 0.5$ eV. Using (20) as the initial guess, we calculate the charge distribution (12) across the entire barrier region, and then calculate the new potential (8) corresponding to the new charge distribution, the electric field as the gradient of $V(z)$, and the corresponding potential due to the polarization (10). We then repeat the whole procedure until convergence (reaching self-consistency (19) between $V(z)$, $\rho(z)$, $\mathcal{E}(z)$, and $P(z)$). The self-consistency between the charge redistribution and the electrostatic potential determines the Schottky barrier, including both its height and thickness (depletion and inversion layer widths).

### III. COMPUTATIONAL APPROACH

#### A. Technical details

The calculations are carried out using density functional theory (DFT) with Perdew–Burke–Ernzerhof (PBE) exchange-correlation functional [68] and projector augmented wave method (PAW) [69],[70] implemented in Quantum Espresso package.[71] For the relaxation of the structures we used the Vienna Ab initio simulation package (VASP) [72],[73] and generalized gradient approximation (GGA) with optB88 exchange-correlation functional including van der Waals correction[74,75,76] in order to account the interaction of Gr with the surface of GaAs. An energy cutoff of 500 eV is used for the plane-wave expansion. The DOS of GaAs crystal is calculated for the primitive cell consisting of two atoms with 31×31×31 **k**-mesh. The interface system is represented by Gr layer in a contact with GaAs(111), which consists of surface layer and five additional layers of GaAs bulk. For surface layer we use surface reconstruction (2x2) with vacancy of Ga, which is described in detail in Section IV.C. The last layer of GaAs is terminated by H.75 pseudo hydrogen atoms in order to make all dangling bonds filled with the electrons from hydrogen atoms. The interface calculation for GaAs-Gr system is taking for 15x15 **k**-mesh, which corresponds to the total of 27 **k**-points. The CBS calculation were performed in cell consisting of three atoms of Ga and three atoms of As which forms three layers of GaAs in order to simulate the decaying process in direction perpendicular to (111) surface. The **k**-mesh of the cell is set to 15×15×15, and the CBS decaying rates were calculated for 27 **k**-points corresponding to the interface calculations in order to calculate the MIGS density in (17).

Si-Al heterostructure was created by contacting Si(100) surface with Al(100) surface in a such way that Al atoms are placed against the hollow positions on Si(100) surface (Fig. 4). The cell consists of twelve Si atoms, six Al atoms, and one hydrogen, which terminates the opposite surface of Si. The lattice parameter of Si crystal is set to 5.430 Å, and Al atoms on the surface (100) are connected to corresponding hollow positions on Si surface. The lattice parameter of Al is decreased from 4.095 Å to 3.840 Å in order to match the Si atom positions, which creates the strain of 6.6% in Al layer. One layer of Si and all Al atoms were allowed to relax in z direction. The Al-Si contact creates chemical bonds between Si and Al atoms. The Si-Si bond length in the contact layer is increased from 2.351 Å to 2.399 Å due to new chemical bonds with Al atoms, and the Al-Al bond length is increased from 2.706 Å in the compressed Al layer to 2.853 Å in the Al-Si contact layer. The relaxed bond length of Si-Al is 2.629 Å. The DOS of Si crystal is calculated for the primitive cell consisting of two atoms, and the **k**-mesh is set to 51×51×51. The interface calculation for Al-Si system is taking for 21×21 **k**-mesh, which corresponds to the total of 121 **k**-points. The CBS calculation were performed in eight atom conventional cell of Si crystal in order to simulate the decaying process in the direction perpendicular to (100) surface. The **k**-mesh of the eight-atom cell is set to 15×15×15, and the CBS decaying rates are calculated for 121 **k**-points corresponding to the interface calculation.

### B. Self-consistent procedure

The self-consistent procedure consists of three independent sets of DFT calculations:

1. DFT calculation of the primitive cell of GaAs for different values of the electric field corresponding to each layer of the semiconductor, to get the DOS $D_b[E]$ of the semiconductor bands, which will be used in (13) and (14), and the induced dipole moments of each layer under the local electric field. In order to get the band gap of GaAs in accordance with experimental value of 1.42 eV we applied the scissor correction.[77,78]

2. Calculation of the complex band structure (CBS) of the GaAs(111) slab surface containing three GaAs layers in order to get the decay rate $\kappa_z$ needed in (17).

3. Calculation of the PDOS $D_i[E,\mathbf{k}]$ of the interfacial layer in Gr-GaAs(111) system. The bottom layer of the Gr-GaAs(111) system is terminated with hydrogen and is not included in the PDOS. PDOS $D_i[E,\mathbf{k}]$ will be used in order to separate light and heavy hole contribution (see Section D) and to calculate MIGS density according to (17).

These DFT calculations are stitched together in the self-consistent procedure by the following additional steps, starting from the initial approximation to the potential $V(z)$ (22) and calculating:

1. the electron and hole charge densities in each layer of the semiconductor (13) and (14) as functions of z using $D_b[E]$, updated from DFT and $V(z)$;

2. the induced charge density due to MIGS using $D_i[E,\mathbf{k}]$, updated from DFT, $V(z)$, and $\kappa_z$ also updated from DFT;

3. the total charge density $\rho(z)$ in (12);

4. the new potential $V(z)$ in (6) and (8) from the total induced charge density $\rho(z)$ and the local electric field $\mathcal{E}(z)$, which will be used in the DFT calculations of the next iteration;

5. the compensating surface charge density σ in (7) and the energy shift of the surface states $\Delta E$ defined in (18).

Steps 2 - 5 are repeated until consistency between $\rho(z)$ and $V(z)$ is reached.

### C. Integration over k-space

DOS of the interface at the gap of the semiconductor has a Gaussian-like distribution at Γ with a finite width. CBS decay rate $\kappa_z(E,\mathbf{k})$ has the slowest decaying rates at Γ point. Because of that at large distances the interfacial DOS $D_i[\sigma,E,\mathbf{k}]\exp[-2\kappa_z(E,\mathbf{k})z]$ will have a sharp peak at Γ, and the numerical integration in (17) over $d\mathbf{k}$ will be inaccurate and produce overestimated values. In order to improve the accuracy of the integration for large distances, we use parabolic approximation for CBS, and write the DOS as a Gaussian distribution over **k**,

$$D_i[E,\mathbf{k},z] \approx D_i[E,0,z]\exp\left[-\frac{k_r^2}{\kappa_z(E,0)}z\right] \qquad (23)$$

where $k_r^2 = k_x^2 + k_y^2$. This allows to calculate the integral (17) over Brillouin zone analytically and connect to the numerical solution at small z by matching the corresponding width of the Gaussian and numerical distributions. The detailed derivation of the integration scheme and the method to connect the numerical and analytical solutions are described in Appendix A.

### D. Light and heavy holes separation in PDOS

PDOS of the interfacial layer contains the contributions from heavy and light holes, which have different decaying rates $\kappa_z(E,\mathbf{k})$. The interfacial layer selected for the PDOS should be far enough from the surface layer, in order not to include the surface states, which are significantly differ from the bulk states. This layer also should be far enough from the bottom layer of the interface calculation, in order not to include the hydrogen states. From six layers of the system for the interface calculation we are taking PDOS for the 3d and 4th layers and writing the expressions,

$$D_i^3[E,\mathbf{k}] = D_i^L[E,\mathbf{k}]\exp\left[-2\kappa_z^L(E,\mathbf{k})z_3\right] + \\ + D_i^H[E,\mathbf{k}]\exp\left[-2\kappa_z^H(E,\mathbf{k})z_3\right]\} \qquad (24)$$

$$D_i^4[E,\mathbf{k}] = D_i^L[E,\mathbf{k}]\exp\left[-2\kappa_z^L(E,\mathbf{k})z_4\right] +$$
$$+ D_i^H[E,\mathbf{k}]\exp\left[-2\kappa_z^H(E,\mathbf{k})z_4\right]\} \quad (25)$$

where $D_i^{3/4}[E,\mathbf{k}]$ are the PDOS calculated for 3d and 4th layers, and $z_{3/4}$ are the positions of the 3$^d$ and 4$^{th}$ layers. Knowing the decaying rates for light and heavy holes, we can solve the set of equations (24) and (25) for $D_i^L[E,\mathbf{k}]$ and $D_i^H[E,\mathbf{k}]$ for each points (E,**k**), and separate the light hole contribution to PDOS from heavy holes in order to use with the corresponding decay rates.

## IV. RESULTS

We now apply our method on a heterostructure of Si(100) surface in a contact with Al(100) surface, and 2D semi-metal in contact with SC,[79] as a GaAs-graphene heterostructure. This two heterostructures have different connection type between metal and semiconductor. For GaAs-graphene system the connection is due to Van der Waals forces, whereas the Si/Al system creates chemical bonds between metal and semiconductor atoms.

The Schottky contact parameters, Schottky barrier height, Schottky potential energy, depletion layer width (DLW), inversion layer width (ILW), surface charge density, and electric field at the surface, depending on doping concentration are summarized at Tables 1 and 2 for Si/Al heterostructure, and Tables 3 and 4 for GaAs-Gr system. Tables 5 and 6 show the dependence of Schottky parameters on temperature. We define the DLW for p-(n-) type doped SC as a region where the main carrier concentration $\rho_p$ ($\rho_n$) together with the background charge $\rho_d$ is reduced by e value from the value at the surface (see Fig. 5 and 6). Correspondingly, we define the ILW for p-(n-) type doped SC as a region where the thermally activated minority n-(p-) type carrier concentration is reduced by value of e from the value at the surface. We should point out here that the extracting the width of depletion layer from linear dependence of electric field due to Schottky potential give close result for DLW obtained from our definition. Thus for GaAs-Gr system for n-type doping of 1.15×10$^{17}$ cm$^{-3}$ the DLW calculated from linear dependence of electric field gives 971 Å, whereas from our method 1,014 Å.

### A. Schottky contact Al(100)-Si(100)

Tables 1 and 2 show the results of self-consistent solutions for SBH, depletion layer width, the inversion layer width, total charge accumulated at the metal surface, and electric field close to the semiconductor surface for the system for n-type and p-type doping, correspondingly. The space distributions for the electrons and holes for depletion and inversion layers are presented in Fig. 5 and 6.

There are many experimental studies on Al-Si contacts using C-V, I-V, and photoelectric measurements[80,81,82,83,84,85] for n- and p-type doped Si. In real experiments the surface of Si often has an oxide layer, which contains positive charge and increase SBH. Card[83] pointed out that the results for SBH is very sensitive to the condition of the Si surface before Al evaporation. Thus, experimental measurements for SBH for n-type doped Si vary from 0.61 to 0.77 eV and depend on growing condition, treatment procedure for oxide layer, annealing of the surfaces, and even on resting time after the treatment. Pellegrini[86,87] pointed out that for n-type semiconductor the SBH derived from C-V measurements should always be higher (and for p-type should always be lower) than those obtained from I-V and photoelectric measurements because quantum-mechanical penetration of electrons from the metal into the semiconductor energy gap makes the barrier lower than that deduced from the capacitance measurements. Thanailakis[84] used I-V measurements and got the values of 0.61 eV for SBH for Al-Si contact prepared by Al evaporation on vacuum cleaved Si surface. Tejedor et.al.[44] calculated with Green function formalism[88] the SBH of 0.60 eV for Al-Si contact by including the surface barrier effect using the jellium model, and surface (virtual) states using a two-band narrow-gap model[89]. Our result for SBH obtained self-consistently for n-type doped Si is 0.59 eV (see Table 1), in good agreement with experimental result of Thanailakis[84] and calculated result by Tejedor et.al.[44]

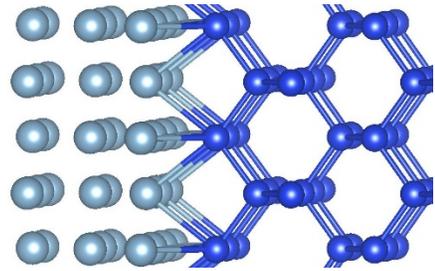

Fig. 4. Al(100)-Si(100) heterostructure. Al atoms in light blue and Si atoms in blue. In the contact layer Si-Al bonds are created.

The depletion layer for doping concentration of 1.72×10$^{19}$ cm$^{-3}$ is 67 Å, whereas the inversion layer is 0.8 Å. The charge accumulated at the metal surface is 1.09×10$^{13}$ cm$^{-2}$, and electric field at the SC surface is 1.66×10$^{-2}$ V/Å. For doping concentration of 1.99×10$^{17}$ cm$^{-3}$ corresponding to one experiment,[83] the electric field at the surface is 1.58×10$^{-3}$ V/Å which is in good agreement with experimental value of 10$^{-3}$ V/Å.[83] Reducing bulk doping (by tuning the Fermi energy) causes the width of depletion and inversion layers to increase, and the charge accumulated at the metal surface to decrease, accompanied by smaller band bending. The smaller charge on the interface results smaller $\Delta E$, and larger $\Phi_B$ (see Fig. 3). This trend is consistent with previous empirical calculation by Osvald[90]. Our results show the SBH of 0.590 eV for low doping concentration and 0.567 eV for higher doping concentration (see Table 1). Experimental measurements for depletion layer width for Si contacts under zero bias shows value from a fraction of a micrometer to tens of micrometers depending on doping concentration and the geometry of the contact.[91]

Inversion layer width (see Tables 1, 2, 5, and 6) depends on doping concentration. For high doping concentration ~$10^{19}$ cm$^{-3}$ the ILW is less than 1 Å, whereas for doping of ~$10^{13}$ cm$^{-3}$ the ILW will be more than 1,000 Å. Many experimental[92,93,94,95,96,97] and theoretical[92,98,99] works show strong influence of inversion layer on Schottky parameters of the structures. Experimentally the inversion layer can be detected for lightly doped semiconductors only.[98] Empirical models[92,98,99] for describing inversion layer are based on the assumption that inversion layer is the interfacial layer of two to three atomic dimensions, and even set the inversion layer width to zero in the charge-sheet approximation.[100] Our results show that for small doping concentration the inversion layer width may exceed 1,000 Å (see Figs. 5, 6, 11, and 12).

Table 1. Schottky barrier height (SBH), potential energy ($-eV$), energy shift $\Delta E$, depletion layer width (DLW), inversion layer width (ILW), surface charge density ($\sigma$), and electric field ($\mathcal{E}$) at the surface as functions of electron doping concentration ($\rho_n$) for Si-Al system for $T$ = 300 K.

| $E_F-E_{VBM}$ (eV) | $\rho_n$ (cm$^{-3}$) | SBH (eV) | $-eV$ (eV) | $\Delta E$ (eV) | DLW (Å) | ILW (Å) | $\sigma$ (cm$^{-2}$) | $\mathcal{E}$ (V/Å) |
|---|---|---|---|---|---|---|---|---|
| 0.78 | -4.02×10$^{13}$ | 0.590 | 0.250 | 4.50×10$^{-5}$ | 27,892. | 1,001. | -1.05×10$^{10}$ | -1.59×10$^{-5}$ |
| 0.89 | -2.72×10$^{15}$ | 0.590 | 0.360 | 4.63×10$^{-4}$ | 4,031. | 87.4 | -1.08×10$^{11}$ | -1.64×10$^{-4}$ |
| 0.93 | -1.28×10$^{16}$ | 0.589 | 0.365 | 1.06×10$^{-3}$ | 1,971. | 37.4 | -2.47×10$^{11}$ | -3.75×10$^{-4}$ |
| 1.00 | -1.99×10$^{17}$ | 0.586 | 0.466 | 4.33×10$^{-3}$ | 553.3 | 8.5 | -1.04×10$^{12}$ | -1.58×10$^{-3}$ |
| 1.12 | -1.72×10$^{19}$ | 0.555 | 0.555 | 3.57×10$^{-2}$ | 66.8 | 0.81 | -1.09×10$^{13}$ | -1.66×10$^{-2}$ |

Table 2. Schottky barrier height (SBH), potential energy ($-eV$), energy shift $\Delta E$, depletion layer width (DLW), inversion layer width (ILW), surface charge density ($\sigma$), and electric field ($\mathcal{E}$) at the surface as functions of hole doping concentration ($\rho_p$) for Si-Al system for $T$ = 300 K.

| $E_F-E_{VBM}$ (eV) | $\rho_p$ (cm$^{-3}$) | SBH (eV) | $-eV$ (eV) | $\Delta E$ (eV) | DLW (Å) | ILW (Å) | $\sigma$ (cm$^{-2}$) | $\mathcal{E}$ (V/Å) |
|---|---|---|---|---|---|---|---|---|
| 0.37 | 1.70×10$^{13}$ | 0.529 | -0.159 | -4.37×10$^{-4}$ | 33,655. | 2,287. | 1.01×10$^{11}$ | 1.19×10$^{-4}$ |
| 0.25 | 1.76×10$^{15}$ | 0.528 | -0.278 | -1.52×10$^{-3}$ | 4,494. | 125.1 | 3.56×10$^{11}$ | 4.42×10$^{-4}$ |
| 0.12 | 2.68×10$^{17}$ | 0.523 | -0.403 | -6.88×10$^{-3}$ | 443.9 | 8.2 | 1.69×10$^{12}$ | 2.38×10$^{-3}$ |
| 0.00 | 2.21×10$^{19}$ | 0.490 | -0.490 | -4.01×10$^{-2}$ | 55.5 | 0.76 | 1.25×10$^{13}$ | 1.87×10$^{-2}$ |

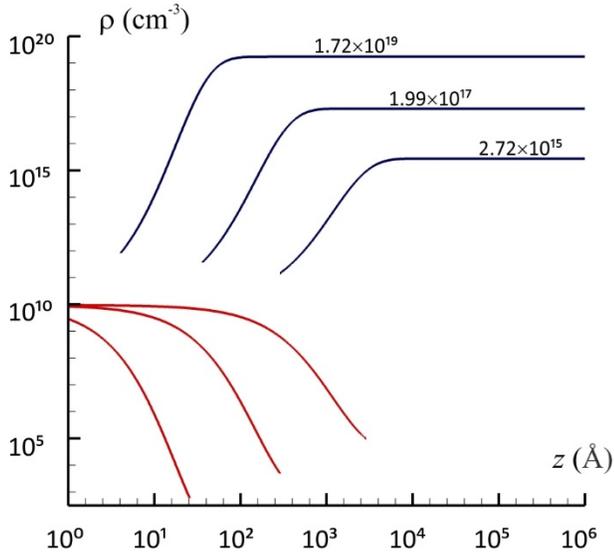

Figure 5: Induced charge density of electrons (absolute value) (blue) and holes (red) for Si-Al system for $T$ = 300 K, $n$-type doping with concentrations $\rho_n$ = 1.72×10$^{19}$ cm$^{-3}$, 1.99×10$^{17}$ cm$^{-3}$, and 2.72×10$^{15}$ cm$^{-3}$, respectively.

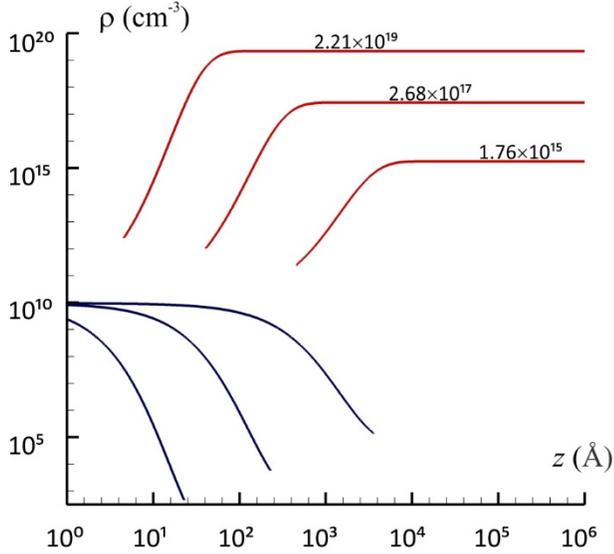

Figure 6: Induced charge density of holes (red) and electrons (absolute value) (blue) for Si-Al system for $T$ = 300 K, $p$-type doping with concentrations $\rho_p$ = 2.21×10$^{19}$ cm$^{-3}$, 2.68×10$^{17}$ cm$^{-3}$, and 1.76×10$^{15}$ cm$^{-3}$, respectively.

### B. Complex band structure of bulk GaAs

Fig. 7 shows the CBS of bulk GaAs along the direction perpendicular to (111) calculated by PWCOND software of Quantum Espresso package.[101,71] The main band $\kappa_z$ responsible for the evanescent states in the gap is bolded in red in Fig. 7 and corresponds to light holes contribution. The smallest penetration rate of MIGS into the bulk of GaAs is near the middle of the gap (midpoint)[9] with the characteristic length of about 20 Å. The penetration rate (decay rate) is increasing (decreasing) as the energy approaches VBM/CBM, and diverges at the CBM/VBM ($\kappa_z = 0$).

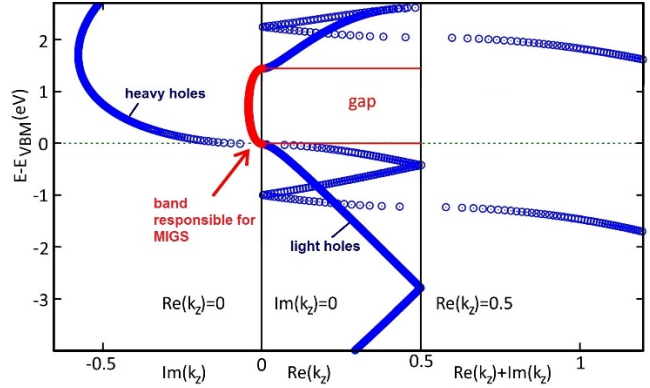

Figure 7: CBS $\kappa_z(E,\mathbf{k})$ of GaAs for (111) surface for Γ point. The band highlighted by red (the slowest decaying band) corresponds to light holes and is responsible for evanescent states in the gap of the semiconductor.

### C. Surface of intrinsic GaAs

When the GaAs semiconductor with $sp^3$ hybridization in the bulk is cleaved to create a surface, the dangling $sp^3$ bonds tend to form a surface with metallic properties.[102] The surface atoms tend to reconstruct in order to reduce the dangling bonds and minimize the surface energy.[103,104,105] GaAs (111) surface is known to have several surface configurations.[106] One of the configurations with minimum surface energy is a (2x2) surface reconstruction with vacancies of Ga atom.[107,108,109] In this configuration on every (2×2) surface unit one Ga atom is removed (see Fig. 8 and 9). Ga vacancies in surface layer create additional dangling bonds on As atoms (dashed red lines in Fig. 8), which form pairs with dangling bonds of Ga atoms exposed to vacuum[103] (see Fig. 8). Such a reconstruction favors relaxation of Ga atoms towards As atoms to form a planar surface with Ga-As $sp^2$ hybridized bonds.[107] Thus, after the reconstruction the metallic character of the surface containing dangling bonds is changed back to a semiconducting character with all bonds filled. Such surface reconstruction creates surface states in the valence band and in the conduction band[110,111,112,113] which are partially filled (see Fig. 3) with the surface acceptor states at 0.27 eV above VBM.[114]

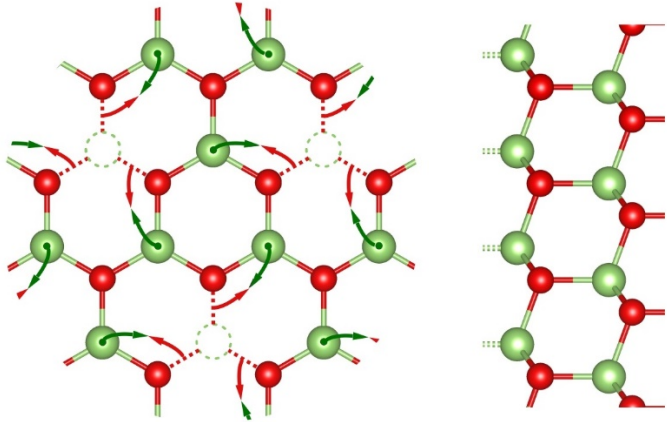

Figure 8: Surface reconstruction (2×2) with vacancy of Ga (shown as dashed green circles) top view on the left, and cleaved surface of GaAs(111) side view on the right. The dangling bonds of Ga atoms exposed to vacuum are shown as green dots on the left and as dashed green lines on the right. The dangling bonds of As atoms are shown by dashed red lines. Dangling bonds of As-Ga tend to combine to eliminate all dangling bonds of the surface. The relaxed surface is shown on Fig. 9.

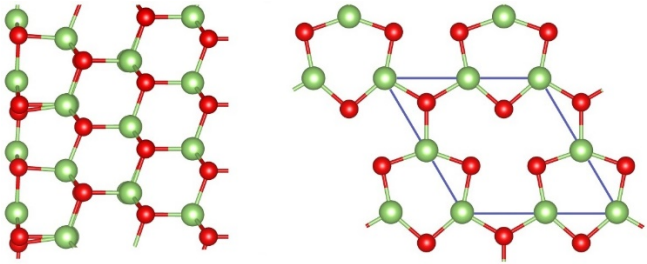

Figure 9: Relaxed flat surface GaAs (111) with (2×2) surface reconstruction with vacancy of Ga. Ga atoms in green, and As atoms in red. The left is the view along *z* direction perpendicular to the surface, and the right is the top view on the surface layer. Each (2×2) unit (shown as a blue quadrangle) has a vacancy of Ga.

We calculated the PDOS of the surface layer and the internal layers of the system containing six layers of GaAs. For intrinsic (undoped) GaAs the Fermi level in the bulk is 0.78 eV (at 300 K) measured from the VBM, as determined from total charge neutrality requirement. The Fermi level and intrinsic concentration of holes and electrons, which compensate each other in undoped GaAs, depends on the temperature.[115]

On the surface, thermally activated surface holes in the surface valence band and electrons in the surface conduction band compensate each other to create the charge neutrality level (CNL)[7,9] on the surface. For undoped GaAs(111) the CNL coincides with the Fermi level of the bulk.[107] Thus this surface is kept uncharged. However some of the semiconducting surfaces may have slightly bent bands.[5]

### D. Surface of doped GaAs

Doped semiconductors can be described simply by the location of the Fermi energy in the gap, which in turn determines the concentration of the carriers. By placing the Fermi energy at a desired energy in the calculation, we can adjust the doping level without dealing with the actual dopants. For example, for a *n*-type doped semiconductor with a clean surface (i.e., not in contact with a metal or an overlayer), the surface layer exhibits electron accumulation filling the surface states, and the layers closer to the bulk have electrons depleted. This causes both the conduction and the valence bands to bend. The band bending moves the band gap away from the Fermi energy, affecting on thermally activated holes and electrons.

In the bulk, doping, for example, with Si atoms as dopants substituting Ga atoms creates excess mobile electrons on $Si_{Ga}$ defects which can freely move over the crystal. The typical resulting doping concentration with Si dopants is up to $10^{18} - 10^{20}$ cm$^{-3}$.[116] The remaining body of positively charged defects $Si_{Ga}^+$ creates a compensating background charge, maintaining charge neutrality in the bulk.

### E. Surface of GaAs(111) in contact with graphene

Fig. 10 shows the relaxed structure of GaAs(111)-Gr contact. Ga and As atoms are placed against the hollow positions of Gr rings. This structure corresponds to one of the possible minimum energy configurations.[117] The strain in the Gr layer is 6.3%. The structure were optimized in VASP using exchange-correlation functional including van der Waals correction. The relaxed distance between the surface of GaAs and the flat Gr layer is 3.32 Å.

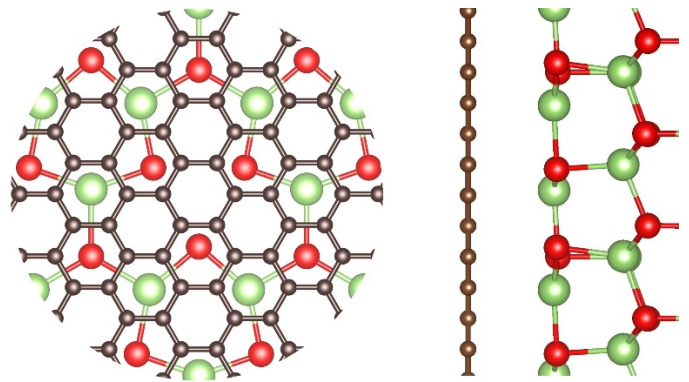

Figure 10: GaAs(111) surface layer in contact with Gr, top view on the left (shown only Gr layer and surface layer of GaAs), Ga atoms in green, As atoms in red, and C atoms in grey. Ga and As atoms are placed against the hollow positions of Gr rings. On the right is side view. The distance between the surface layer of GaAs and the Gr layer is 3.32 Å.

Tables 3 and 4 shows the calculated SBH, the depletion layer width, the inversion layer width, the accumulated charge at the surface, and the electric field near the SC surface depending on the concentration for $n$- (Table 3) and $p$-type (Table 4) doping. The depletion layer appears due to band bending of the conduction band, whereas the inversion layer appears due to band bending of the valence band (see Fig. 3).

Table 3. Schottky barrier height (SBH), potential energy (-$eV$), energy shift $\Delta E$, depletion layer width (DLW), inversion layer width (ILW), surface charge density ($\sigma$), and electric field ($\mathcal{E}$) at the surface as functions of electron concentration ($\rho_n$) for GaAs-Gr system for $T$ = 300 K.

| $E_F$-$E_{VBM}$ (eV) | $\rho_n$ (cm$^{-3}$) | SBH (eV) | -$eV$ (eV) | $\Delta E$ (eV) | DLW (Å) | ILW (Å) | $\sigma$ (cm$^{-2}$) | $\mathcal{E}$ (V/Å) |
|---|---|---|---|---|---|---|---|---|
| 1.19 | -2.00×10$^{13}$ | 0.738 | 0.508 | 3.43×10$^{-5}$ | 63,770. | 884. | -1.21×10$^{10}$ | -1.49×10$^{-5}$ |
| 1.31 | -2.07×10$^{15}$ | 0.738 | 0.628 | 3.82×10$^{-4}$ | 6,995. | 74.0 | -1.37×10$^{11}$ | -1.70×10$^{-4}$ |
| 1.40 | -5.94×10$^{16}$ | 0.736 | 0.716 | 2.16×10$^{-3}$ | 1,395. | 12.8 | -7.89×10$^{11}$ | -9.75×10$^{-4}$ |
| 1.42 | -1.15×10$^{17}$ | 0.735 | 0.735 | 3.01×10$^{-3}$ | 1,014. | 9.2 | -1.11×10$^{12}$ | -1.37×10$^{-3}$ |

Table 4. Schottky barrier height (SBH), potential energy (-$eV$), energy shift $\Delta E$, depletion layer width (DLW), inversion layer width (ILW), surface charge density ($\sigma$), and electric field ($\mathcal{E}$) at the surface as functions of hole concentration ($\rho_p$) for GaAs-Gr system for $T$ = 300 K.

| $E_F$-$E_{VBM}$ (eV) | $\rho_p$ (cm$^{-3}$) | SBH (eV) | -$eV$ (eV) | $\Delta E$ (eV) | DLW (Å) | ILW (Å) | $\sigma$ (cm$^{-2}$) | $\mathcal{E}$ (V/Å) |
|---|---|---|---|---|---|---|---|---|
| 0.32 | 1.05×10$^{14}$ | 0.680 | -0.360 | -1.19×10$^{-3}$ | 23,270. | 494.5 | 4.32×10$^{11}$ | 4.41×10$^{-4}$ |
| 0.26 | 1.07×10$^{15}$ | 0.680 | -0.420 | -1.60×10$^{-3}$ | 7,879. | 136.8 | 5.82×10$^{11}$ | 6.07×10$^{-4}$ |
| 0.20 | 1.09×10$^{16}$ | 0.679 | -0.479 | -2.36×10$^{-3}$ | 2,646. | 39.3 | 8.65×10$^{11}$ | 9.38×10$^{-4}$ |
| 0.14 | 1.11×10$^{17}$ | 0.677 | -0.537 | -4.29×10$^{-3}$ | 880.2 | 11.3 | 1.60×10$^{12}$ | 1.83×10$^{-3}$ |
| 0.08 | 1.12×10$^{18}$ | 0.672 | -0.592 | -9.80×10$^{-3}$ | 293.3 | 3.3 | 3.84×10$^{12}$ | 4.58×10$^{-3}$ |
| 0.00 | 2.00×10$^{19}$ | 0.650 | -0.650 | -3.20×10$^{-2}$ | 73.6 | 0.73 | 1.48×10$^{13}$ | 1.81×10$^{-2}$ |

Table 5. Schottky barrier height (SBH), potential energy (-$eV$), energy shift $\Delta E$, depletion layer width (DLW), inversion layer width (ILW), surface charge density ($\sigma$), and electric field ($\mathcal{E}$) at the surface as functions of the temperature for electron doping corresponding to Fermi level at CBM for GaAs-Gr system.

| T (K) | $\rho_n$ (cm$^{-3}$) | SBH (eV) | -$eV$ (eV) | $\Delta E$ (eV) | DLW (Å) | ILW (Å) | $\sigma$ (cm$^{-2}$) | $\mathcal{E}$ (V/Å) |
|---|---|---|---|---|---|---|---|---|
| 150 | -3.72×10$^{16}$ | 0.737 | 0.737 | 1.73×10$^{-3}$ | 1,797. | - | -6.42×10$^{11}$ | -7.93×10$^{-4}$ |
| 200 | -5.89×10$^{16}$ | 0.736 | 0.736 | 2.18×10$^{-3}$ | 1,421. | 7.7 | -8.04×10$^{11}$ | -9.93×10$^{-4}$ |
| 250 | -8.46×10$^{16}$ | 0.736 | 0.736 | 2.60×10$^{-3}$ | 1,188. | 8.5 | -9.59×10$^{11}$ | -1.18×10$^{-3}$ |
| 300 | -1.15×10$^{17}$ | 0.735 | 0.735 | 3.01×10$^{-3}$ | 1,014. | 9.2 | -1.11×10$^{12}$ | -1.37×10$^{-3}$ |
| 350 | -1.49×10$^{17}$ | 0.735 | 0.735 | 3.42×10$^{-3}$ | 889.9 | 9.8 | -1.26×10$^{12}$ | -1.56×10$^{-3}$ |

Table 6. Schottky barrier height (SBH), potential energy (-$eV$), energy shift $\Delta E$, depletion layer width (DLW), inversion layer width (ILW), surface charge density ($\sigma$), and electric field ($\mathcal{E}$) at the surface as functions of the temperature for hole doping corresponding to Fermi level at VBM for GaAs-Gr system.

| T (K) | $\rho_p$ (cm$^{-3}$) | SBH (eV) | -$eV$ (eV) | $\Delta E$ (eV) | DLW (Å) | ILW (Å) | $\sigma$ (cm$^{-2}$) | $\mathcal{E}$ (V/Å) |
|---|---|---|---|---|---|---|---|---|
| 150 | 5.81×10$^{18}$ | 0.664 | -0.664 | -1.79×10$^{-2}$ | 137.8 | - | 8.49×10$^{12}$ | 1.03×10$^{-2}$ |
| 200 | 9.82×10$^{18}$ | 0.659 | -0.659 | -2.28×10$^{-2}$ | 105.8 | 0.62 | 1.07×10$^{13}$ | 1.30×10$^{-2}$ |
| 250 | 1.46×10$^{19}$ | 0.654 | -0.654 | -2.75×10$^{-2}$ | 86.6 | 0.68 | 1.28×10$^{13}$ | 1.56×10$^{-2}$ |
| 300 | 2.00×10$^{19}$ | 0.650 | -0.650 | -3.20×10$^{-2}$ | 73.6 | 0.73 | 1.48×10$^{13}$ | 1.81×10$^{-2}$ |
| 350 | 2.61×10$^{19}$ | 0.645 | -0.645 | -3.63×10$^{-2}$ | 64.4 | 0.78 | 1.67×10$^{13}$ | 2.05×10$^{-2}$ |

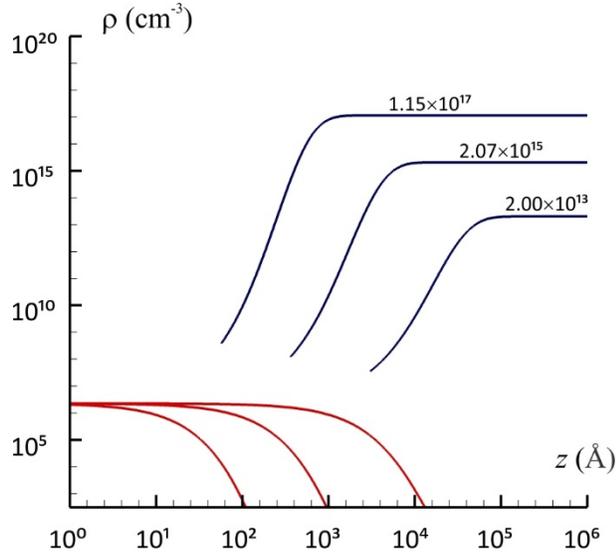

Figure 11: Self-consistent solution for induced charge density of electrons (absolute value) (blue) and holes (red) for $T$ = 300 K, $n$-type doping with concentrations $\rho_n$ = 1.15×10$^{17}$ cm$^{-3}$, 2.07×10$^{15}$ cm$^{-3}$, and 2.00×10$^{13}$ cm$^{-3}$, respectively.

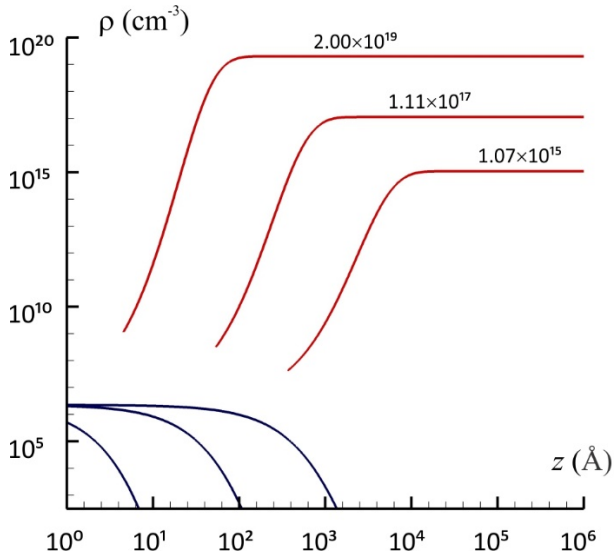

Figure 12: Self-consistent solution for induced charge density of holes (red) and electrons (absolute value) (blue) for $T$ = 300 K, $p$-type doping with concentrations $\rho_p$ = 2.00×10$^{19}$ cm$^{-3}$, 1.11×10$^{17}$ cm$^{-3}$, and 1.07×10$^{15}$ cm$^{-3}$, respectively.

The characteristics of Schottky contact are sensitive to many parameters like doping concentration and temperature. The self-consistent solution for the majority and minority carrier concentrations are also shown on Fig. 11 for $n$-type and on Fig. 12 for $p$-type doping, correspondingly. For Fermi level at CBM, the effective concentration of $n$-type doping is 1.15×10$^{17}$ cm$^{-3}$ and the depletion layer is 1,014 Å, whereas the inversion layer is 9.2 Å. The charge accumulated at the surface is 1.03×10$^{12}$ cm$^{-2}$, and the Schottky barrier height is 0.737 eV. Decreasing Fermi level position or decreasing the concentration of $n$-type doping, the surface charge is also decreased, and the widths DLW and ILW are increased. The SBH is slightly decreased by 0.001 eV. For $p$-type doping concentration of 2.00×10$^{19}$ cm$^{-3}$, which corresponds to Fermi level at VBM, the DLW is 74 Å and the ILW is 0.7 Å, whereas the SBH is 0.661 eV. Lowering the concentration of $p$-type doping (increasing Fermi level), the SBH is increased and the widths DLW and ILW are also increased. The SBH (Tables 3 and 4) are listed for $n$- and $p$-type doping, correspondingly.

The space charge layer width for GaAs experimentally measured by photothermal radiometric deep level transient spectroscopy ranges 10-100 μm depending on the treatment of the sample.[114] Calculated values of the depletion layer width (Tables 3 and 4) are from below 100 to over 10,000 Å depending on doping concentration.

Tables 5 and 6 show the dependence of SBH, DLW, ILW, σ, and electric field near the SC surface on temperature for $n$-type and $p$–type doping, respectively. Increasing temperature reduces the depletion and inversion layer widths due to increasing carrier concentration. This behavior shows qualitative agreement with experimental data for Au-GaAs measured by $I$-$V$ curves.[50]

## V. CONCLUSION

We have developed a first-principles theory for calculating Schottky barrier height and the widths of depletion and inversion layers without using any empirical parameters. The theory is combining self-consistently the DFT calculations of the interface and the bulk layers under the influence of the electric filed due to Schottky potential, with the Poisson equation for the excess electrostatic potential due to charge redistribution on Schottky contact. The method is tested on the GaAs(111)-graphene and Si(100)/Al(100) complexes. The depletion layer arises due to band bending of conduction band, whereas the inversion layer arises due to band bending of valence band. The SBH is a consequence emerging from the self-consistency between the redistributed charges in the bulk of the semiconductor due to MIGS, and thermally activated holes and electrons, and the electrostatic potential arising from these induced charges. We find that the Schottky barrier height, the depletion layer width, and the inversion layer width depend on many factors including temperature, and doping concentration. The Schottky barrier height is decreasing with increasing the doping concentration as a consequence of filling of the surface states. We did not consider local inhomogeneities on the contact surfaces or local changes of the SBH due to doping atoms on the surface. The effects of electron tunneling through Schottky barrier and thermionic field emission are also not included in our theory. Our method can have potential impact on modeling junctions where a Schottky barrier is present and allow one to compare experiments with first-principles calculations. This capability can be used to guide computational design of heterogeneous junctions. Further studies of GaAs-Gr with defects in graphene layer and adsorbed atoms, as well a magnetic molecule-graphene-GaAs complex are underway.

## ACKNOWLEDGEMENTS

This work was supported as part of the Center for Molecular Magnetic Quantum Materials (M2QM), an Energy Frontier Research Center funded by the U.S. Department of Energy (DOE), Office of Science, Basic Energy Sciences under Award DE-SC0019330. This research used resources of the National Energy Research Scientific Computing Center (NERSC), a U.S. DOE Office of Science User Facility operated under Contract No. DE-AC02-05CH11231.

## APPENDIX A. INTEGRATION OVER K-SPACE

In parabolic approximation [118] the CBS decaying wave vector depends on $k_x$ and $k_y$ in the form,

$$\kappa_z(E, k_r) = \sqrt{\kappa_z^2(E, 0) + k_r^2} \quad (26)$$

where $k_r^2 = k_x^2 + k_y^2$.

At large distance the PDOS is sharply peaked near $k_r = 0$, thus we can approximate,

$$\kappa_z(E, k_r) \approx \kappa_z(E, 0) + \frac{k_r^2}{2\kappa_z(E, 0)} \quad (27)$$

This allows us to approximate the integrand in (17) by the Gaussian distribution

$$\exp\left(-\frac{k^2}{\delta^2}\right) \quad (28)$$

with a small width δ,

$$D_i^G[E, k_r, z] \approx D_i[E, 0, z] \exp\left[-\frac{k_r^2}{\kappa_z(E, 0)} z\right] \quad (29)$$

where

$$D_i[E, 0, z] = D_i[E, 0] \exp\left[-2\kappa_z(E, 0) z\right] \quad (30)$$

is the DOS of the interface distribution at $z$.
Using (29) the integration over $d\mathbf{k}$ in (17) gives,

$$D_{\text{MIGS}}^G(E, z) = D_i[E, 0] \exp\left[-2\kappa_z(E, 0) z\right] \times \frac{1}{(2\pi/a)^2} \frac{\kappa_z(E, 0)}{z} \left[1 - \exp\left(-\frac{(2\pi/a)^2}{\kappa_z(E, 0)} z\right)\right] \quad (31)$$

where $a$ is the lattice parameter.
The connection between numerical integration over $d\mathbf{k}$ (17) and the analytical integration (31) can be made by comparison of the width of the Gaussian distribution for numerical and analytical solutions. The numerical solution is valid at small $z$, whereas the Gaussian solution is valid for $z$ far from the interface. The Gaussian parameter $\delta^2$ of (29) is

$$\delta^2 = \frac{\kappa_z(E, 0)}{z} \quad (32)$$

whereas the parameter $\delta^2$ of the numerical solution (17) can be calculated numerically by several $k$-points close to Γ. Calculating the integrand in (17) at point $k_*$ and using (28) we can write the Gaussian parameter $\delta^2$ for the numerical solution at the distance $z$ in the form,

$$\delta^2 = \frac{k_*^2}{2z[\kappa_z(E, k_*) - \kappa_z(E, 0)] - \ln(D_i[E, k_*]/D_i[E, 0])} \quad (33)$$

The pure Gaussian distribution (32) has a parameter $\delta^2$ diverging at $z = 0$, whereas $\delta^2$ for the real distribution (33) has a finite value at $z = 0$ due to the Gaussian type distribution $D_i[E, \mathbf{k}]$ near Γ with a finite width at the interface.
Compare (33) with (32) allows us to write the distance $z^*$ matching the two solutions,

$$z^* = \frac{-\ln(D_i[E, k_*]/D_i[E, 0])}{k_*^2/\kappa_z(E, 0) - 2[\kappa_z(E, k_*) - \kappa_z(E, 0)]} \quad (34)$$

---